\documentstyle[multicol,aps,epsf,rotate,amsfonts,amssymb,prb]{revtex}
\newcommand{\D}[2]{\frac{\partial #1}{\partial #2}}
\begin{document}

\draft

\title{Electron-phonon renormalization of the absorption edge of the
  cuprous halides}

\author{J. Serrano$^{\rm a}$, Ch. Schweitzer$^{\rm b}$,
C.~T.~Lin$^{\rm a}$, K. Reimann$^{\rm c}$, M. Cardona$^{\rm a}$, and
D.~Fr\"ohlich$^{\rm b}$}

\address{$^{\rm a}$Max-Planck-Institut f\"ur Festk\"orperforschung,
  Heisenbergstr.~1, 70569 Stuttgart, Germany\\
  $^{\rm b}$Institut f\"ur Physik,
  Universit\"at Dortmund, 44221 Dortmund, Germany\\
  $^{\rm c}$Max-Born-Institut f\"ur Nichtlineare Optik und Kurzzeitspektroskopie,
  12489 Berlin, Germany
  }

\date{\today}

\maketitle

\begin{abstract}
Compared to most tetrahedral semiconductors, the temperature dependence of the
absorption edges of the cuprous halides (CuCl, CuBr, CuI) is very small. CuCl
and CuBr show a small increase of the gap $E_0$ with increasing temperature,
with a change in the slope of $E_0$ vs. $T$ at around 150~K: 
above this temperature, the variation of
$E_0$ with $T$ becomes even smaller. This unusual behavior has been clarified
for CuCl by measurements of the low temperature gap vs. the isotopic masses of
both constituents, yielding an anomalous negative shift with increasing copper
mass. Here we report the isotope effects of Cu and Br on the gap of CuBr, and
that of Cu on the gap of CuI.  The measured isotope effects allow us to
understand the corresponding temperature dependences, which we also report, to
our knowledge for the first time, in the case of CuI.  These results enable us
to develop a more quantitative understanding of the phenomena mentioned for
the three halides, and to interpret other anomalies reported for the
temperature dependence of the absorption gap in copper and silver
chalcogenides; similarities to the behavior observed for the copper
chalcopyrites are also pointed out.
\end{abstract}

\pacs{PACS: 71.35.Cc,
  71.36.+c,
  63.20.Kr 
  }

\begin{multicols}{2}

\section{Introduction}
\label{sec:intro}

The electronic band structure of the cuprous halides differs considerably from
that of other tetrahedral semiconductors which contain neither copper nor
silver as a cation.  It was early recognized that the differences are due to a
strong admixture of the $3 d^{10}$ copper electrons ($4 d^{10}$ in the case of
silver) to the conventional, valence-band-forming $p$ electrons of the anion
($2 p^7$ for chlorine). Instead of the conventional 8 valence electrons of the
primitive cell of zincblende ($s^2 + p^6$) a total of 18 ($d^{10} + s^2 +
p^6$) hybridized electron states constitute the valence bands of the cuprous
halides. The ten $d$ electron states split in the cubic field into six
$\Gamma_{15}$ and four $\Gamma_{12}$ states. The former hybridize strongly
with the six $p$-like valence states of the halogen atoms, whereas the
$\Gamma_{12}$ states do not hybridize at ${\bf k} = 0$, but they hybridize
with $s$ and $p$ states at a general point of the Brillouin zone (BZ)
(continuity requires this hybridization to remain small).
This hybridization leads to a number of
anomalies, several of them have been discovered after the original suggestion
in Ref.~\onlinecite{Car63} (for an early review see Ref.~\onlinecite{Gol77}).
Among them, we first mention that the lowest gap of these materials (which is
direct and occurs at the $\Gamma$ point, as in GaAs and ZnSe) is lower than
expected from extrapolation of their isoelectronic group IV, III-V and II-VI
counterparts (Ge: 0.9~eV, GaAs: 1.5~eV, ZnSe: 2.7~eV, CuBr: 3.0~eV). Even more
striking is the fact that the spin-orbit splitting $\Delta_0$ of the uppermost
valence band ($\Gamma_{15}$ splits into $\Gamma_8$ and $\Gamma_7$ double group
representations) is rather small, even negative ($\Gamma_7$ above $\Gamma_8$)
for CuCl.\cite{Car63,Shi65} Additionally, anomalously small
temperature\cite{Car63} and pressure\cite{Ves81,Bla83,RE94A} dependences of
the gap have also been observed.

The temperature dependence of semiconductor gaps is intrinsically connected to
their dependence on isotopic mass. The reason for this is that the temperature
dependence is caused, to a large part, by the renormalization of the
band gap due to electron-phonon interaction.
Because of the zero-point vibrations there is
already an effect at zero temperature, which is proportional to the
square of the vibrational amplitudes, obviously mass-dependent.

During the past ten years a number of
experiments\cite{Goe98,Etch92,Par94,Car01a,Las00,Goe99} has been performed on
the dependence of the gap on isotopic mass. In compound semiconductors, such
experiments yield considerably more information than temperature-dependent
experiments, since the mass of each constituent element can have a different
effect on a given gap.\cite{Goe98,Goe99} Of particular interest is the case of
CuCl: an increase of the chlorine mass (from $^{35}$Cl to $^{37}$Cl) leads to
the usual increase of the $E_0$ gap, whereas the increase of the copper mass
(from $^{63}$Cu to $^{65}$Cu) results in an anomalous decrease of the
gap.\cite{Goe98} This led to the understanding of the anomalous temperature
dependence of the $E_0$ gap of CuCl reported in
Refs.~\onlinecite{Rag67,Kai71,Mil84,Lew81}. Using the same model and the
observed data for $E_0(T)$ in CuBr, the authors of Ref.~\onlinecite{Goe98}
predicted the dependences of the $E_0$ gap on copper ($M_{\rm Cu}$) and
on bromine mass ($M_{\rm Br}$).

In this paper we report measurements of $\partial E_0/\partial M_{\rm Cu}$ and
$\partial E_0/\partial M_{\rm Br}$ for CuBr, which confirm the above
conjecture. We also report measurements of $\partial E_0/\partial M_{\rm Cu}$
for the $E_0$ gap of CuI (since iodine is one of the few elements with only
one stable isotope, $^{127}$I, $\partial E_0/\partial M_{\rm I}$ could not be
measured). For CuI we also present rather unusual results on the temperature
dependence of the $E_0$ gap. In spite of the missing value of $\partial
E_0/\partial M_{\rm I}$ for CuI, still a detailed analysis of electron-phonon
renormalization effects is possible. This is due to the weak and considerably
structured temperature dependence, as reported in this work.
In the temperature range from 0 to 300~K
the $E_0$ gap decreases by 8~meV, in agreement with the early data of
Ref.~\onlinecite{Car63}. A fit to the experimental $E_0(T)$
data with a three-oscillator model 
yields values for $\partial E_0/\partial M_{\rm I}$ and for $\partial
E_0/\partial M_{\rm Cu}$, the latter value in good agreement with the directly
measured one. The results we obtained for the cuprous halides are used to
interpret the anomalies reported for the temperature dependence of the lowest
band gap in other tetrahedral compounds like cuprous and silver
chalcopyrites,\cite{Chi95,Qui92,Wei98,Riv97,Qui89,Art87,Ali86,Qui90} which
exhibit a small positive slope at low temperatures followed by a sign reversal
in the slope above $\approx 100$~K.

This article is organized as follows:  We discuss in the next section the
peculiar band structure of the cuprous halides in the region around the $E_0$
gap. Section~\ref{sec:exp} contains details about the experimental method. In
Sec.~\ref{sec:results} we present our data for the isotopic dependence of
$E_0$ at low temperature for CuI and CuBr and for the temperature dependence
of $E_0$ for CuI. Then we introduce in Sec.~\ref{sec:discussion} 
a model for both temperature and isotope
effects and compare our present results with those obtained in
Ref.~\onlinecite{Goe98} for CuCl. Finally, in Sec.~\ref{sec:chalco},
an exhaustive compilation of the
anomalies reported  in the behavior of the temperature dependence of $E_0$ is
performed for the cuprous and silver chalcopyrites and quantitatively
interpreted in terms of our results for the simpler cuprous halides.

\section{Electronic Band Structure}
\label{sec:theo}

The lowest absorption edge (equivalently referred to as lowest
 gap or lowest exciton) of CuCl differs considerably from that of CuBr and
CuI. The difference is due to the anomalous sign of the spin-orbit splitting,
which is negative for CuCl, but positive for the other cuprous
halides,\cite{Car63,Shi65,Goe98} a consequence of the negative contribution of
the copper $3d$ electrons to the spin-orbit splitting.\cite{Shi65} Hence the
symmetry of the uppermost valence band of CuBr and CuI ($E_0$) is $\Gamma_8$,
whereas that of CuCl is $\Gamma_7$ ($E_0 + \Delta_0$). The symmetry of the
lowest conduction band is $\Gamma_6$ in all cases.

Because of Coulomb interaction, electrons in the conduction band and holes in
the valence band form exciton states. Excitons corresponding to
the $\Gamma_8$ valence band are often labeled $Z_{1,2}$, where $Z$ stands for
zincblende, whereas excitons belonging to the $\Gamma_7$ valence band are
labeled $Z_3$. The exciton symmetry is given by the direct product of
conduction and valence band symmetries. Thus $Z_{1,2}$ excitons ($\Gamma_8
\times \Gamma_6$) are eightfold degenerate and split through exchange
interaction into $\Gamma_3$, $\Gamma_4$, and $\Gamma_5$
components,\cite{Sug76} whereas the $Z_3$ fourfold degenerate
excitons ($\Gamma_7 \times \Gamma_6$),
split into $\Gamma_2$ and $\Gamma_5$. Only
the $\Gamma_5$ components are dipole allowed and contribute to one-photon
absorption processes. Two-photon transitions, on the other hand, are allowed
for $\Gamma_1$, $\Gamma_3$, $\Gamma_4$ and $\Gamma_5$ excitons.
Each of the $\Gamma_5$
excitons constitutes a triplet, whose transverse components interact strongly
with photons, thus giving rise to exciton-photon coupled states (polaritons).

\section{Experiment}
\label{sec:exp}

We have performed measurements on samples of CuI and CuBr with all possible
combinations of stable isotopes ($^{63}$Cu, $^{65}$Cu; $^{127}$I; $^{79}$Br,
$^{81}$Br) and the corresponding natural compositions. Details of the crystal
growth have been published elsewhere.\cite{Ser01,Lin96,Lin01}

To be able to accurately determine the rather small isotope shifts, we have
used the nonlinear optical technique of two-photon absorption\cite{FR94A}
(TPA), since this technique yields much smaller line widths than linear
optical methods, as, e.\,g., reflection or absorption (as an example, for CuCl
one can compare the absorption data in Ref.~\onlinecite{Car63} with the
two-photon data in Refs.~\onlinecite{Goe98} and~\onlinecite{FR71A}). For the
same reason we have restricted our measurements to the excitons with the
lowest energies, i.\,e., to the $Z_{1,2}$ excitons in CuBr and CuI, since
excitons with higher energies are broadened through phonon-induced decay into
lower-energy states.

The classical method\cite{FR71A} for the measurement of TPA is to detect a
change in the intensity of the transmitted light of the first beam during the
presence of the second (high-intensity) beam.  This method is, however, not
sensitive enough to detect TPA in our thin samples. Therefore in our case TPA
is detected by monitoring the luminescence intensity from a defect-related
level below the band gap as a function of the sum of the two photon energies
(the two-photon energy). Since we use two photons from the same
laser, the two-photon energy is equal to twice the laser photon energy.

The experimental setup for these measurements consists of an exciting laser, a
cryostat with the sample and the detection system.

Since non-resonant in the intermediate state
(the photon energies are far from resonances in our
experiments) two-photon absorption in these compounds is a very weak effect;
high excitation intensities of about 10~MW/cm$^2$ must be used.  They are
generated by a Nd:YAG-laser-pumped tunable dye (styril) laser with a pulse
width of 5~ns and a repetition rate of 10~Hz.

The samples were mounted strain-free in a helium flow cryostat. The
temperature, which is measured with a calibrated silicon diode mounted near
the samples, can be varied by changing the rate of helium flow and 
by additional heating. 
For the measurements of the isotopic effect on the band gap the
samples were held at a temperature of 7~K.

The detection system consists of a $f/1.8$ collection optics, a prism
spectrometer and appropriate optical filters to separate the weak luminescence
from the intense laser light, and a bialkali photomultiplier. The wavelength
of the spectrometer is held fixed at the luminescence maximum. The output
signal from the photomultiplier is fed into a gated integrator and sent via an
analog-digital converter to a personal computer, which also controls the
scanning of the dye laser.

\section{Results}
\label{sec:results}

\subsection{CuI}
\label{sec:cui}

Two-photon absorption spectra for CuI with different copper isotopic
compositions are shown in Fig.~\ref{fig1}. The lower and higher energy
peaks we observe are
identified as the $\Gamma_3$ (with possible admixture of $\Gamma_5$
transverse) and $\Gamma_5$ longitudinal excitons.\cite{Sug76,Sug80} The
splitting of 6.7~meV between the peaks is in agreement with previously
reported values\cite{Sug80} (6.5~meV). Shown in Fig.~\ref{fig2} are the peak
energies obtained from a Lorentzian fit vs. the average isotopic mass. The
values of $\partial E_0/\partial M_{\rm Cu}$ from a linear least-squares fit
are listed in Table~\ref{tab1}. Although the error bars of the $\Gamma_3$
data, $-550 \pm 12$~$\mu$eV/amu, are much smaller than those of the weaker
$\Gamma_5$ data, $-510 \pm 150$~$\mu$eV/amu, the agreement between both sets
of data is excellent.

Figure~\ref{fig3} shows the measured temperature dependence of the
lower-frequency peak of Fig.~\ref{fig1} for natural CuI for temperatures up to
300~K. Measurements at higher temperature are not meaningful because of the
increasing line width (see error bars in Fig.~\ref{fig3}). One should note the
following important points:

\begin{enumerate}

\item The overall effect of temperature is a decrease in energy, opposite to
the observations\cite{Goe98} for CuCl and CuBr.

\item At about 150~K a change in the slope of $E_0(T)$ is seen, remindful
  of the behavior observed for CuCl and CuBr.

\item Below 60~K a weak sigmoidal behavior is observed, with maxima at
$T = 0$ and $T = 40$~K (see inset in Fig.~\ref{fig3}).
\end{enumerate}

\subsection{CuBr}
\label{sec:cubr}

Some two-photon spectra for CuBr samples of different isotopic compositions
measured at 7~K are displayed in Fig.~\ref{fig4}. In contrast to CuI here one
observes only the $\Gamma_5$ longitudinal exciton. 
Attempts to observe the $\Gamma_3$ exciton revealed that it is at least two
orders of magnitude weaker than in CuI. This may be related to the near
degeneracy of the $\Gamma_3$ and $\Gamma_5$ excitons in CuI, which does not
occur\cite{Sug80,Noz82} in CuBr.
By means of Lorentzian fits
we obtain the $E_0$ energies plotted in Fig.~\ref{fig5} as a function of the
corresponding isotopic masses of copper and bromine. Least-squares fits to
these points result in the corresponding slopes listed in Table~\ref{tab1}.
The data allow us to establish the anomalous negative sign of
the copper mass derivative which was predicted in Ref.~\onlinecite{Goe98}.

\section{discussion}
\label{sec:discussion}

We assume here, as it was done in previous\cite{Car63,Goe98} works, that
shifts with either temperature or isotopic mass observed for the $Z_{1,2}$
excitons (or for $Z_3$ in the case of CuCl) are representative of the
corresponding one-electron gaps. This assumption is usually justified by the
fact that the exciton binding energies, which are very large in the copper
halides, are nevertheless more than one order of magnitude smaller than the
corresponding band gap. Additionally, one has to consider that for CuCl (and
to a lesser extent also for CuBr) the exciton binding energy\cite{LB1} 
($\approx 190$~meV for CuCl,
$\approx 140$~meV for CuBr) is close to or larger than
another important characteristic energy, that of the spin-orbit splitting
$\Delta_0$ ($\approx -70$~meV for CuCl, $\approx 150$~meV for CuBr).
Consequently the spin-orbit splitting measured for the exciton should be
somewhat smaller than the splitting $\Delta_0$ of the valence bands at the
$\Gamma$ point of the BZ.  This decrease in $\Delta_0$, which has been
investigated for diamond,\cite{Ser00,Car01b} has not been discussed for CuCl
and CuBr. It is nevertheless unlikely that changes in $\Delta_0$ induced by
changes either of temperature or of mass will alter our conclusions. This has
been confirmed for the temperature dependence\cite{Lew81} of $E_0$ and $E_0 +
\Delta_0$ in CuBr (see Fig.~8 of Ref.~\onlinecite{Goe98}). In any case, for
CuI the problem does not arise since\cite{Car63} $\Delta_0\approx 630$~meV is
much larger than the exciton binding energy\cite{Got68} ($\approx 58$~meV).

The model we use here for the description of the mass and temperature
dependences of the band gap is the same as in Ref.~\onlinecite{Goe98}. The
effects of both temperature and mass changes can be divided into the effect
of the change of lattice constant and the renormalization by the
electron-phonon interaction. The change in lattice constant leads to the
following change of the gap:

\begin{equation}
\Delta_{\rm th}E_0(X) = 3\alpha\, \frac{a(X) - a_0}{a_0}, \label{eq:1}
\end{equation}
with $\alpha = dE_0/d \ln V$ the deformation potential of the $E_0$ gap and
$a(X)$ the lattice constant as a function of $X$ (temperature or one of the
isotopic masses).

In principle, for the calculation of the renormalization by
electron-phonon interaction it is necessary to include every possible phonon
mode. It turns out, however, that a much simplified model is sufficient for a
good description of the data. Instead of using the complete phonon dispersion,
we use only two (three for CuI) average phonons. This leads to the following
expression:

 \begin{equation}
   \Delta E(T) = \sum_i\frac{A_i}{\Omega_i M_i} \left[ 2\,n_B\left(\frac{\Omega_i}{T}\right) +
   1\right].
   \label{eq:2}
 \end{equation}
The sum is over the oscillators used, $A_i$ is the electron-phonon coupling
coefficient for the oscillator $i$, $\Omega_i$ its frequency (in Kelvin), and
$n_B$ the corresponding Bose-Einstein factor. 
$M_i$ is the reduced mass for the vibration
considered.

Two oscillators with opposite values of $A$ are required to describe the
change in $E_0(T)$ observed at $\approx 100$~K for both CuCl and CuBr. The
anomalous positive slope for $T< 100$~K is due to the copper vibrations
($A_{\rm Cu} > 0$), whereas the decrease in slope for $T> 100$~K is due to the
vibrations of the halogen, which have negative values of $A$.

Since it is not clear {\em a priori} which phonons should be used, we compare
in Table~\ref{tab1} different possibilities. 
We show in this table the experimental
results for the isotope dependence together with theoretical values obtained
from a two-oscillator fit of the temperature dependence. The three values
given in Table~\ref{tab1} for CuCl and CuBr are obtained by choosing for the
two oscillators (T) the average TO and the average TA phonon energy, (L) the
LO and the LA phonon energy, or (M) the average of all acoustic phonon
energies and the average of all optical phonon energies, respectively. 
As one can see, the
best agreement between measured and calculated values is reached for
CuCl for case T (the same energies were also used in Ref.~\onlinecite{Goe98})
and for CuBr for case M. At this point we should note that in
Ref.~\onlinecite{Goe98} copper, in spite of its slightly lighter mass,
seemed to play the main role in the acoustic vibrations, whereas bromine was
responsible for the optical modes. This {\em ad hoc} assumption has recently
been validated by Raman measurements\cite{Ser01} on isotopically tailored
CuBr. 
A similar assumption is also necessary to explain the present measurements.

In order to get an appropriate description of the temperature 
dependence of the gap of CuI shown 
in Fig.~\ref{fig3}, the use of a three-oscillator model is necessary. 
For the frequencies
we choose the average TA, LA, and LO phonon energies. The results of this fit,
together with the effect of thermal expansion, are shown by the dashed line in
Fig.~\ref{fig3}. Apart from an overestimate of the 
amplitude of the low-temperature
wiggle, the fit agrees well with the experimental data. Using the model, we
obtain a copper isotopic shift of $-524~\mu$eV/amu, which agrees perfectly
with the measured values ($-550$ and $-510~\mu$eV/amu).

One should note that in CuI the contribution to $E_0(T)$ due to thermal
expansion,\cite{note1} which is shown by the diamonds in Fig.~\ref{fig3}, is
very important\cite{note1} (it accounts for about two thirds of the total band
gap change). This is in marked contrast to CuCl and CuBr, for which 
the effect of the thermal expansion is negligible.

Figure~\ref{fig3} indicates that the net effect of the electron-phonon
interaction (dotted line in Fig.~\ref{fig3}) is very small ($\leq 5$~meV).
This suggests that there is near total  compensation between the effects of
the copper and the iodine vibrations.  The residual effects contain detailed
structure, which requires three oscillators for its description. 
We have placed these three
oscillators at the average TA frequency (60~K), LA frequency (156~K) and
LO frequency (224~K). An additional oscillator (e.\,g., at the TO
frequency) does not improve the quality of the fit. As adjustable parameters
we use the corresponding three coefficients $A_i/(\Omega_i M_i)$
of Eq.~(\ref{eq:2}).
In this manner we obtain, after adding the thermal expansion effect, the dashed
line of Fig.~\ref{fig3}.  The coefficient $A_i/\Omega_i M_i$ that corresponds
to the TA phonons is found to be about one tenth of that of the other two
phonons. Because of the heavier mass of iodine, the LA coefficient should be
proportional to $\partial E_0/\partial M_{\rm I}$, according to
Eq.~\ref{eq:2}, whereas the LO coefficient should correspond to $\partial
E_0/\partial M_{\rm Cu}$, neglecting the TA coefficient. The corresponding
values of $\partial E_0/\partial M_{\rm Cu}$ and $\partial E_0/\partial M_{\rm
I}$ are listed in Table~\ref{tab1}. The fitted value of $\partial E_0/\partial
M_{\rm Cu} = -524\pm 16~\mu$eV/amu (the error bars have been estimated from the
value of the TA coefficient, which was neglected when determining $\partial
E_0/\partial M_{\rm Cu}$) agrees very well with the directly measured one
($-550\pm 12~\mu$eV/amu).

\section{Chalcopyrites}
\label{sec:chalco}

Information about the temperature dependence of the lowest gaps (usually
direct and also labeled $E_0$) of the copper and silver chalcopyrites
(I-III-VI$_2$ compounds) is found scattered throughout the literature,
mostly limited to
the  0 -- 300~K region.\cite{Chi95,Qui92,Wei98,Riv97,Qui89,Art87,Ali86,Qui90}
Most of these data show anomalies similar to those observed for the cuprous
halides, a fact which seems to have remained unnoticed by most authors (note,
however, that in Ref.~\onlinecite{Qui92} {\em ad hoc} fits with
two oscillators, with contributions of opposite signs, were performed). 
We have listed
in Table~\ref{tab2} the gaps of a number of such chalcopyrites and the change
they experience from 0~K to 300~K, together with the temperature $T_a$ at
which a change in the slope of $E_0(T)$ is observed. We have also included in
the last column the corresponding deformation potentials  $\alpha = d E_0/d
\ln V$, taken from Refs.~\onlinecite{Wei98} and~\onlinecite{LB1}. 
For comparison we have also
shown the available corresponding values for the cuprous halides, for
wurtzite structure AgI (AgCl and AgBr crystallize in the rocksalt structure,
therefore presenting a very different behavior),
 and for several zincblende-type III-V and II-VI
compounds. The trends observed for these parameters in the chalcopyrites
reinforce the arguments made in this article concerning the anomalous sign of
the copper (which also applies to silver) contributions to the 
properties under discussion, to
which we can add now the deformation potential $\alpha$:

\begin{enumerate}
\item The differences $E_0(300~K) - E_0(0)$ reverse sign and become
  smaller in magnitude when going from the III-V and II-VI compounds
  to CuCl and CuBr.
\item For CuI and AgI these differences also become very small although
  the sign reversal does not quite take place.
\item The corresponding behavior for the chalcopyrites lies between that of the
  II-VI and the I-VII compounds.
\end{enumerate}

This last statement is reinforced when following the isoelectronic series:

\begin{tabular}{lcccc}
&GaAs &ZnSe   &CuGaSe$_2$  &CuBr\\
 $E_0(T=0)$ (meV): &1540   &2770   &1720   &2967\\
$E_0(300~K)-E_0(0)$:&-95   &-90   &-35   &+32\\
$\alpha$ (eV): & -9.8  & -4.9  & -3.9  & -0.3
\end{tabular}

The near cancellation of $E_0(300 K)-E_0(0)$ for CuI and its negative sign
can also be followed through an isoelectronic series:\\
\begin{tabular}{lcccc}
&GaSb  &ZnTe   &CuGaTe$_2$  &CuI\\
 $E_0(T=0)$ (meV): &813   &2381   &1000   &3057\\
$E_0(300~K)-E_0(0)$:&-88   &-111   &-43   &-8\\
$\alpha$ (eV): & -7.0  & -5.9  & -3.4  & -1.1
\end{tabular}

Note that whereas in the GaAs series $E_0$ increases with temperature for the
chalcopyrite at low temperature and decreases above $T_a$, for the GaSb series
no increase is observed at low temperature, a fact which corresponds to the
results of Fig.~\ref{fig3} for CuI. The values of $E_0(300~K)-E_0(0)$ for the
chalcopyrites lie between those of the corresponding isoelectronic I-VII and
II-VI compounds; this reflects the fact that only half as much copper (or
silver) is present in the chalcopyrites as in the I-VII counterparts. Similar
trends to those just described also apply to the deformation potential
$\alpha$. Hence an anomalous, positive contribution to $\alpha$ must result
from the presence of the noble metal.

We notice that most elements in the chalcopyrites of Table~\ref{tab2} have
more than one stable isotope (exceptions: Al, As and I). It would be most
interesting to measure the corresponding isotope effects on the $E_0$ gaps
of chalcopyrite samples.

\section{Conclusions}
\label{sec:conclus}

We have measured the effect of changing copper isotopes on the two-photon
spectra of CuI and the temperature dependence of the lowest exciton,
corresponding to the $E_0$ gap. For CuBr we have measured the corresponding
isotope effects for both constituents. The results have been used to interpret
the anomalous temperature dependence of the $E_0$ gaps of these materials. The
behavior of $E_0(T)$ is particularly interesting for CuI where we have nearly
complete cancellation of the contributions of the copper and iodine vibrations
to the temperature shift of $E_0$. For CuBr, a reversal of the phonon
eigenvectors was needed to explain the isotope effects, in agreement with
results previously reported from Raman measurements.\cite{Ser01}

We have examined similar effects observed for the copper and silver
chalcopyrites (I-III-VI$_2$) and pointed out that the contribution of the nobel
metals is also negative but less than that in the binary halides.  This follows
naturally from the amount of the noble metal present in these compounds.

\acknowledgments
J. S. acknowledges financial support from the Max-Planck-Gesellschaft
and the Ministerio de Educaci\'on, Cultura y Deportes (Spain) through the Plan
Nacional de Formaci\'on del Profesorado Universitario.

\end{multicols}
\newpage

\begin{minipage}{14cm}
\begin{table}
  \caption{Dependence of the exciton energies on isotope mass and
    corresponding zero-temperature renormalizations.
    The values obtained by varying the isotopes
    of copper and halogen (X) are compared with those found from two-oscillator
    fits to the temperature dependence, the latter labeled $T$, $L$ or $M$
    according to the frequencies chosen for each oscillator, i.e., either
    TA and TO, or LA and LO, or a mixture of tranverse and longitudinal
    phonons of frequencies in the one-third to two-thirds ratio, respectively.
    In the case of CuI, a three-oscillator model was necessary to fit
    the experimental data; we used the average frequencies of the TA,
    LA, and LO bands. The frequencies are given in K, whereas the derivatives
    are displayed in $\mu$eV/amu. $E_0$ is given in meV.}
  \label{tab1}
  \begin{tabular}{lrrrrcrrrrrrr}
    & \multicolumn{4}{c}{Isotope measurements}& & \multicolumn{7}{c}{Temperature data}\\\hline
    &&&&&& $E_0$ & $\D{E^T_0}{M_{\rm Cu}}$
    &$\D{E^T_0}{M_{\rm X}}$& & & &\\[3pt]
    CuX &${\D{E^{\Gamma_3}_0}{M_{\rm Cu}}}$&$\D{E^{\Gamma_3}_0}{M_{\rm X}}$&
    $\D{E^{\Gamma_5}_0}{M_{\rm Cu}}$&$\D{E^{\Gamma_5}_0}{M_{\rm X}}$&&$E_0$&
    $\D{E^L_0}{M_{\rm Cu}}$&$\D{E^L_0}{M_{\rm X}}$&$\omega_{\rm TA}$&
    $\omega_{\rm LA}$&$\omega_{\rm TO}$&$\omega_{\rm LO}$\\[3pt]
    &&&&&&$E_0$&$\D{E^M_0}{M_{\rm Cu}}$&$\D{E^M_0}{M_{\rm X}}$& & & &\\[3pt]\hline\hline
        &     &     &     &     && 3236 &  -69 &  546 &&&&\\
    CuCl& -81$^a$ & 346$^a$ & -71 & 382 && 3260 & -446 & 1482 & 51 & 169 & 300 & 333\\
        &     &     &     &     && 3244 & -152 &  773 &&&&\\\hline
        &     &     &     &     && 2975 &  -44 &   87 &&&&\\
    CuBr& &&-115(90)& 132 (40)&  & 2984 & -378 &  401 & 53 & 158 & 216 & 233\\
        &     &     &     &     && 2978 &  -98 &  142 &&&&\\\hline
    CuI & -550(12)& & -510(150)&&& 3045 & -524 &  243 & 60 & 156 & 192 & 224
\end{tabular}
\tablenotetext{$^a$Because of the different band edge structure of CuCl,
  these data do not correspond to a $\Gamma_3$ exciton but to the upper
  polariton.  See Fig.~\ref{fig1} of Ref.~\onlinecite{Goe98}.}
\end{table}
\end{minipage}

\begin{minipage}{14cm}
\begin{table}
  \caption{Temperature shift of the lower absorption gap, $E_0$, for
    cuprous and silver chalcopyrites. The data are compared with that of the
    corresponding isoelectronic binary compounds, the cuprous halides, and other
    binary compounds with zinc-blende crystal structure. $T_a$ stands for the
    temperature at which a kink is observed in the temperature dependence, and
    the hydrostatic deformation potential is listed under $\alpha$.
    $E_0$ and $E_0(300 K)-E_0(0)$ are given in meV,$\alpha$ in eV,
    and $T_a$ in K.}

  \begin{tabular}{rrrrd}
    & $E_0(T=0)$& $E_0(300 K)-E_0(0)$ & $T_a$ & $\alpha$ \\\hline
    $^a$CuAlS$_2$  & 3510 & -45 & 140 &     \\
    $^b$CuGaS$_2$  & 2510 & -60 & 120 &     \\
    $^b$CuGaSe$_2$ & 1720 & -35 & 140 & -3.9$^c$ \\
    $^d$CuGaTe$_2$ & 1444 & -88 &     & -4.6$^c$ \\
    $^b$CuInS$_2$  & 1534 & -13 & 120 &     \\
    $^b$CuInSe$_2$ &  990 & -23 & 100 & -2.2$^c$ \\
    $^e$CuInTe$_2$ & 1000 & -43 & 200 & -3.4$^c$ \\
    $^f$AgGaS$_2$  & 2710 & -33 & 110 &     \\
    $^f$AgGaSe$_2$ & 1820 & -22 & 170 & -3.2$^c$ \\
    $^g$AgInSe$_2$ & 1200 & +10 & 120& -1.7$^c$ \\
    $^h$AgInTe$_2$ & 1000 & -36 &     &     \\
    $^i$CuCl       & 3206 & +50 & 100 & -0.4 \\
    $^i$CuBr       & 2967 & +32 & 100 & -0.3 \\
    $^i$CuI        & 3057 & -8  & 180 & -1.1 \\
    $^i$AgI        & 2900 & -12 &     &     \\
    $^i$GaAs       & 1540 & -95 &     & -9.8 \\
    $^i$GaSb       &  813 & -88 &     & -7.0 \\
    $^i$ZnSe       & 2770 & -90 &     & -4.9 \\
    $^i$ZnTe       & 2381 &-111 &     & -5.9 \\
  \end{tabular}
  \label{tab2}
  \tablenotetext{$^a$Ref.~\onlinecite{Chi95}}
  \tablenotetext{$^b$Ref.~\onlinecite{Qui92}}
  \tablenotetext{$^c$Ref.~\onlinecite{Wei98}}
  \tablenotetext{$^d$Ref.~\onlinecite{Riv97}}
  \tablenotetext{$^e$Ref.~\onlinecite{Qui89}}
  \tablenotetext{$^f$Ref.~\onlinecite{Art87}}
  \tablenotetext{$^g$Ref.~\onlinecite{Ali86}}
  \tablenotetext{$^h$Ref.~\onlinecite{Qui90}}
  \tablenotetext{$^i$Ref.~\onlinecite{LB1}}
\end{table}
\end{minipage}


\begin{figure}
\caption{Two-photon-absorption spectra of isotopically modified CuI measured
at 7~K.
The two features observed in each spectrum correspond to the
$\Gamma_3$ and the $\Gamma_5$ excitons.
  The dashed
  lines are guides to the eye. Note the anomalous negative shift in the peak
  energies with increasing copper mass.}
\label{fig1}
\end{figure}

\begin{figure}
  \caption{Isotope mass dependence of the $\Gamma_3$ and $\Gamma_5$ excitons in CuI.  The lines
    represent least-squares fits to the experimental data from Fig.~\ref{fig1}.}
  \label{fig2}
\end{figure}

\begin{figure}
\caption{Temperature dependence of the lowest exciton peak for CuI (equivalent
to
  that of the $E_0$ gap).  The dotted line
  represents a three-oscillator fit of the difference between the experimental
  data (solid circles) and the contribution due to the thermal expansion (open
  diamonds), which is displayed shifted by $E_0(T=0)$ in order to show more clearly
  the relevance of this term.
  The dashed line is obtained by adding both contributions. The arrows
  indicate the frequencies of the three oscillators used in the fit (also listed
  in Tab.~\ref{tab1}). A small sigmoidal dependence is observed at low
  temperatures (see inset). Note the flattening of the temperature dependence at
  $\approx 180$~K.}
\label{fig3}
\end{figure}

\begin{figure}
\caption{Two-photon-absorption spectra of CuBr for several isotopic
compositions measured at 7~K.
  The dashed lines are guides to the eye. An anomalous decrease of the exciton
  energy is observed with increasing copper mass (upper figure), whereas the lower
  figure displays the usual positive slope with increasing bromine mass.}
\label{fig4}
\end{figure}

\begin{figure}
\caption{Dependence of the low temperature exciton energy in CuBr on
copper and bromine
  masses. The solid lines represent least-squares fits to the data of
  Fig.~\ref{fig4}.}
\label{fig5}
\end{figure}

\newpage

\begin{minipage}{15cm}
\begin{figure}
\epsfxsize=15cm \centerline{\epsffile{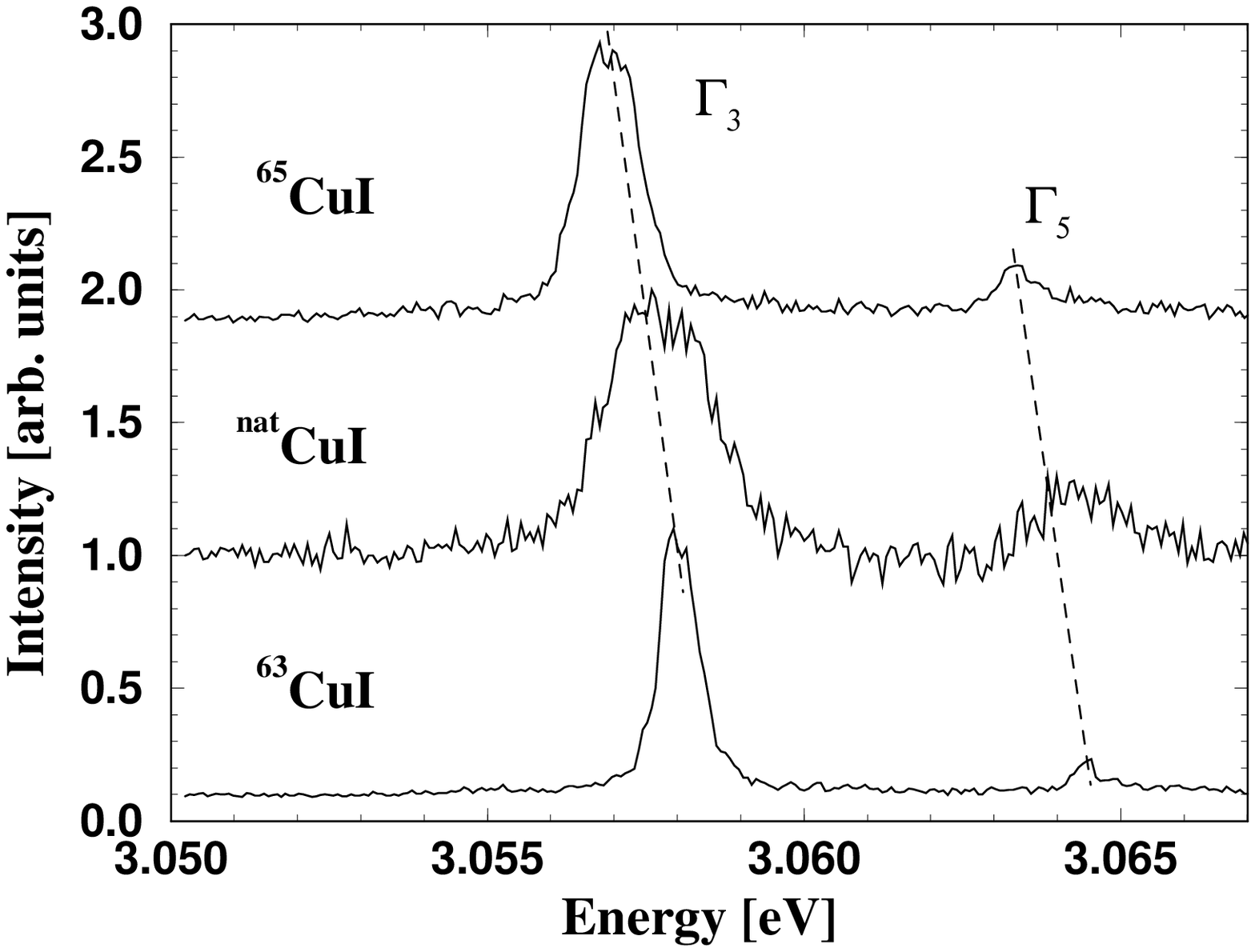}} \vspace*{1cm}
\text{FIG.~\ref{fig1}: J. Serrano {\em et al.}}
\end{figure}
\end{minipage}

\begin{minipage}{15cm}
\begin{figure}
\epsfxsize=15cm \centerline{\epsffile{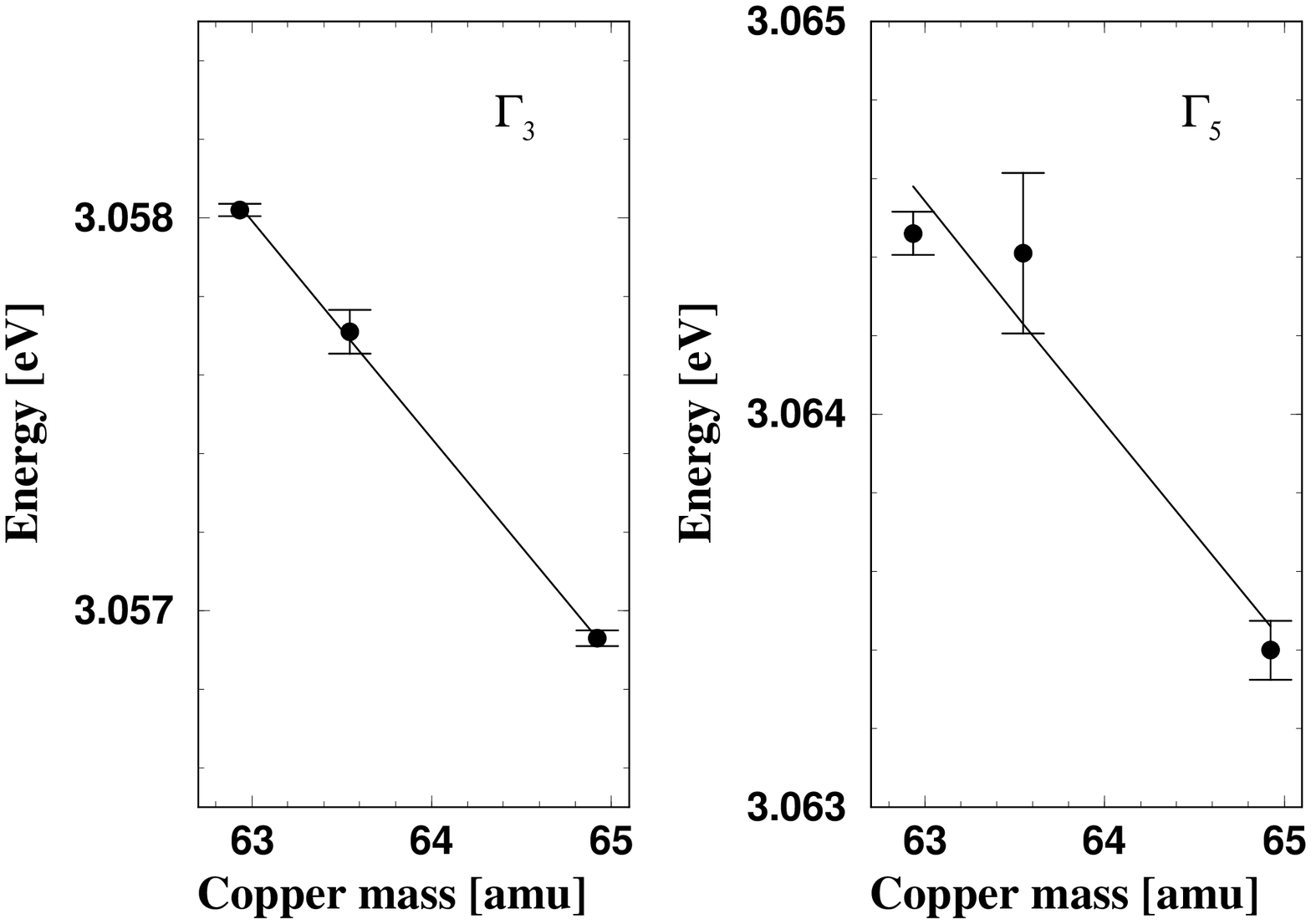}} \vspace*{1cm}
\text{FIG.~\ref{fig2}: J. Serrano {\em et al.}}
\end{figure}
\end{minipage}

\begin{minipage}{15cm}
\begin{figure}
\epsfxsize=15cm \centerline{\epsffile{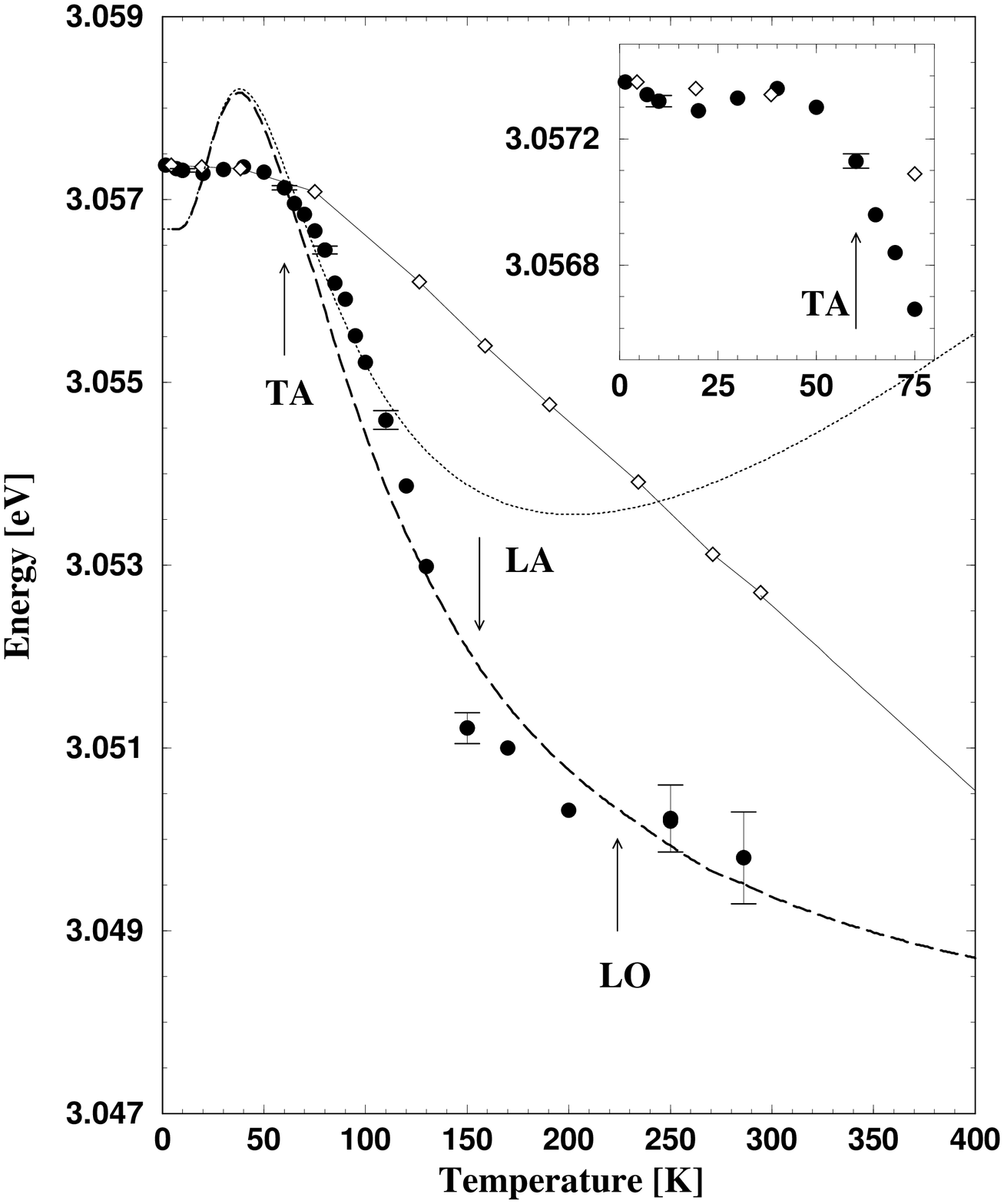}} \vspace*{1cm}
\text{FIG.~\ref{fig3}: J. Serrano {\em et al.}}
\end{figure}
\end{minipage}

\begin{minipage}{15cm}
\begin{figure}
\epsfysize=15cm
\centerline{\epsffile{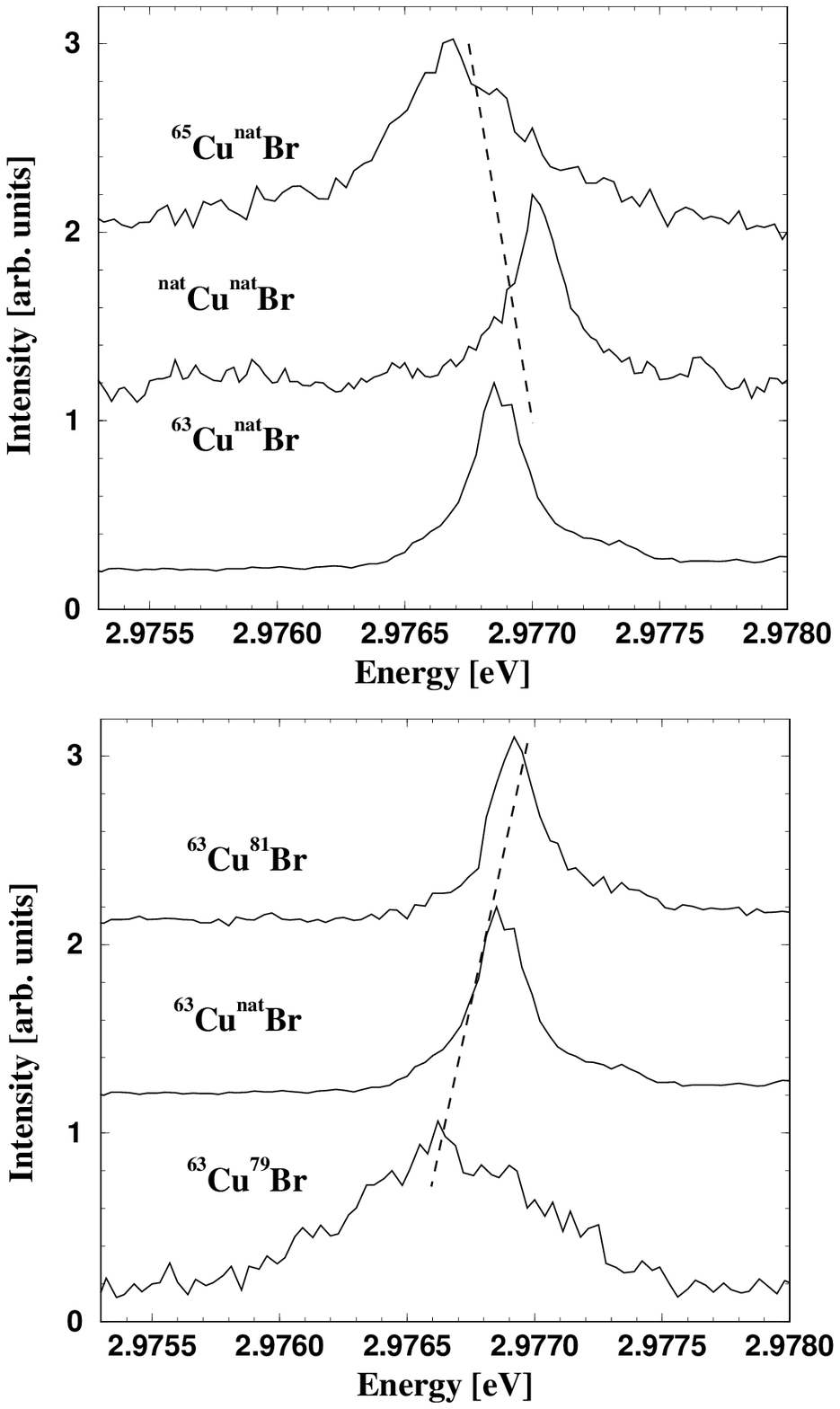}} \vspace*{1cm} \text{FIG.~\ref{fig4}: J.
Serrano {\em et al.}}
\end{figure}
\end{minipage}

\begin{minipage}{15cm}
\begin{figure}
\epsfxsize=15cm \centerline{\epsffile{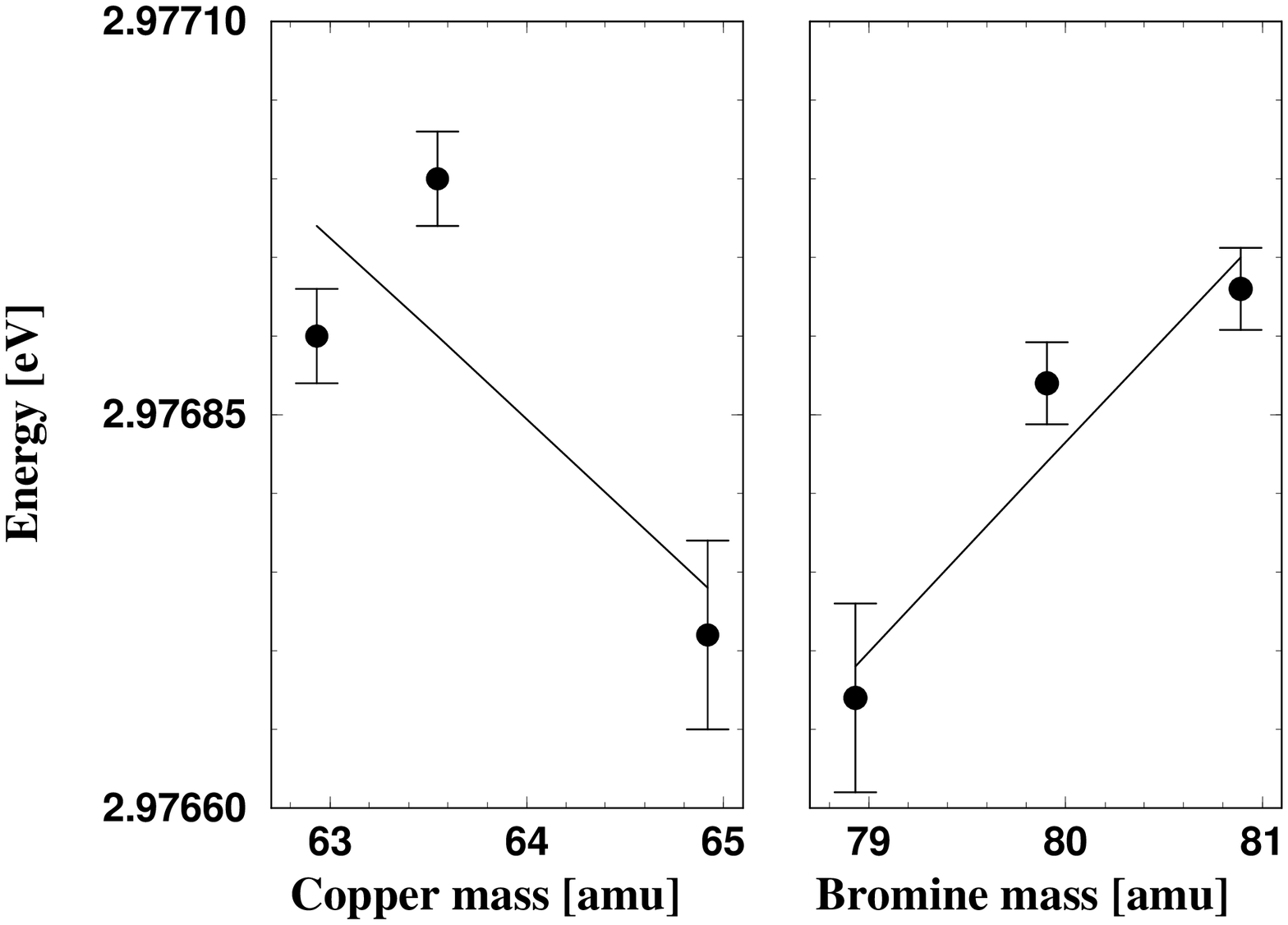}} \vspace*{1cm}
\text{FIG.~\ref{fig5}: J. Serrano {\em et al.}}
\end{figure}
\end{minipage}


\begin{references}
\bibitem{note1}
  We have taken for this calculation $\alpha = -1.2$~eV (-1.10(7)
  from Ref.~\onlinecite{LB1}).
\end{references}
\end{document}